\documentclass[twocolumn,amsmath,amssymb,epsfig,aps,prl]{revtex4}
\usepackage{amsmath}
\usepackage{epsfig}
\usepackage{graphicx}
\usepackage{color}
\usepackage{amssymb}
\usepackage{epstopdf}
\begin{document}
\title{Probing the Efimov discrete scaling in atom-molecule collision}
\author{M. A. Shalchi$^1$}
\author{M. T. Yamashita$^1$}
\author{M. R. Hadizadeh$^{2,3}$}
\author{E.  Garrido$^{4}$}
\author{Lauro Tomio $^{1,5}$} 
\author{T. Frederico$^5$}
\affiliation{$^1$Instituto de F\'{\i}sica Te\'orica, Universidade Estadual Paulista, 01140-070, S\~ao Paulo, SP, Brazil\\
$^2$ Institute for Nuclear and Particle Physics and Department of Physics and Astronomy, 
Ohio University, Athens, OH 45701, USA \\
$^3$ College of Science and Engineering, Central State University,  Wilberforce,
OH 45384, USA \\
$^4$Instituto de Estructura de la Materia, IEM-CSIC, Serrano 123, E-28006 Madrid, Spain
\\
$^5$Instituto Tecnol\'ogico de Aeron\'autica, DCTA, 12228-900, S\~ao
Jos\'e dos Campos, SP, Brazil}

\date{\today}

\begin{abstract}
The discrete Efimov scaling behavior, well-known in the low-energy spectrum of 
three-body bound systems for large scattering lengths (unitary limit), is identified 
in the energy dependence of atom-molecule elastic cross-section in mass imbalanced systems. 
That happens in the collision of a heavy atom with mass $m_H$ with a weakly-bound 
dimer formed by the heavy atom and a lighter one with mass $m_L \ll m_H$.
Approaching the heavy-light unitary limit the $s-$wave elastic cross-section $\sigma$ will present 
a sequence of zeros/minima at collision energies following closely
the Efimov geometrical law. Our results open a new
perspective to detect the discrete scaling behavior from low-energy scattering data, 
which is timely in view of the ongoing experiments with ultra-cold binary mixtures having strong 
mass asymmetries, such as Lithium and Caesium or Lithium and Ytterbium.
\end{abstract}

\maketitle

The Efimov effect~\cite{1970efimov} refers to a discrete scaling symmetry, which emerges in the quantum 
three-body system at the unitary limit (when the two-body scattering lengths diverge). 
The optimal condition to observe this discrete scaling symmetry 
in cold atomic laboratories is  now found for heteronuclear three-atom systems
with large mass asymmetry  and large interspecies scattering lengths.
In the Efimov (unitary) limit, the shallow three-body levels are geometrically spaced, namely 
the ratio between the binding energies of the $n$ and $n+1$ levels is given by $B_3^{(n)}/B_3^{(n+1)}=\exp{(2\pi /s_0)}$, 
where $s_0$ is a universal constant, which depends only on the mass ratio and not on 
the details of the interaction. The energy ratio for three identical bosons is
$\exp{(2\pi /s_0)}\approx 515$, decreasing for the case of two heavy particles and
light one. 
When $m_L/m_H  = 0.01 $, for example, the value of this energy ratio goes to 
$\exp{(2\pi /s_0)}= 4.698$~\cite{2006braaten}.

The Efimov geometric scaling factor has been measured in a cold-atom experiment with mass-imbalance mixtures of 
Caesium ($^{133}$Cs) and Lithium ($^6$Li) gases by different groups~\cite{2014tung,2016ulmanis}.
The ratio between the positions of two successive  peaks in the three-body recombination rate, obtained by varying the 
large negative scattering lengths ($a_{HL}$), was found in close agreement with the theory. 
Complementary to this finding, a fingerprint of the Efimov scaling can be found in the $s-$wave ultra-cold atom-molecule 
cross-section by varying the incident momentum energy $k$ instead of the scattering lengths. Natural, but not yet evidenced 
experimentally or theoretically. What we expect is beyond the trimer crossing the corresponding continuum, which creates 
the resonant enhancement of the inelastic collisions of Caesium atoms with Caesium dimers, as observed by Knoop et 
al.~\cite{2009knoop}. 

Furthermore, there is an evident strong interest in ultra-cold heteronuclear atom-molecule collisions  by experimental 
groups~\cite{2010barontini,2013bloom,2014hu}.  Trap setups with ultra-cold degenerated mixtures of 
alkali-metal-rare-earth molecules with strong mass-imbalanced systems as Ytterbium and Lithium ($^{174,173}$Yb$-^6$Li) 
have also been reported in Refs.~\cite{2011hara,2011hansen}. We should mention that on the theory side~\cite{2015makrides}, 
reactions at ultra-cold temperatures with three-body systems such as $^6$Li +  $^{174}$Yb$^6$Li were also investigated.
Therefore, the present possibilities to manipulate collisions with Lithium(Li)-Caesium(Cs)~\cite{2006krems} and 
Ytterbium-Lithium~\cite{2011hara,2011hansen}, as well as the molecules of LiCs and LiYb in ultra-cold 
experimental setups~\cite{2006kraft}, open new opportunities to probe the discrete Efimov scaling with the large 
mass asymmetries. This can be achieved by using low-energy collisions of a heavy atom, such as Caesium or 
Ytterbium, in the weakly-bound molecules as LiCs or LiYb, with $m_L/m_H=$0.045 and 0.034, respectively.  

Going back in time, what was known theoretically from 
the pioneer works for the tri-nucleon 
systems~\cite{1960delves,1967vanoers,1976whiting,1979girard},  was  the
existence of a pole in the spin doublet $s-$wave neutron-deuteron $k\cot\delta_0$, 
which was associated with a virtual state in the tri-nucleon system. Furthermore, such pole is also 
present in the neutron-$^{19}$C scattering~\cite{2008yamashita,2016shalchi,2017Deltuva}, with a corresponding 
pronounced minimum of the $s-$wave elastic cross-section.  If the Efimov geometrical factor decreases,
which is possible with atomic systems, 
our expectation for $H-(HL)$ collision is that several minima in the $s-$wave elastic cross-sections or 
poles in the $k\cot\delta_0$ would
emerge from the characteristic log-periodic behavior carried by the wave-function in the region where the
 Efimov long-range potential is dominant and being reflected in a geometrical 
law for the spacing of the energies of cross-section  minima corresponding to these  poles.
 
 In this letter, we show that the $s-$wave elastic cross-section for the $H-(HL)$ collision 
has minima at geometrically spaced incident energies, for large  values of $a_{HL}$ near the unitary limit. 
We compute the $s-$wave 
phase-shift using the three-body Faddeev formalism with zero- and short-ranged interactions, as 
well as by considering the Born-Oppenheimer (BO) approximation~\cite{1979fonseca}.  
The real part of the $s-$wave phase shift ($\delta_0$) shows zeros and $k\cot\delta_0$ has a sequence of 
poles at colliding energies which tend to follow the Efimov geometric scaling. 

The BO approximation applied to the $H-(HL)$ system provides 
a universal long-range attractive $1/R^{2}$ effective potential ($R$ is the relative $H-H$ distance) 
close to the unitary limit, which acts up to distances $\sim |a_{HL}|$, as shown in Ref.~\cite{1979fonseca}.
At short distances,  the BO potential brings the details of the finite range pairwise potentials expressed as
a boundary condition at $R_0<<|a_{HL}|$ that determines the reference energy $B_3$. The eigenstates of the $H-H$ effective 
hamiltonian has the characteristic log-periodic solutions for $R_0\lesssim  R \lesssim |a_{HL}|$, which leads
to the geometrical ratio between the binding energies and also to the log-periodic properties of $s-$wave scattering 
observables. We extend the procedure used in ~\cite{1979fonseca} to the
scattering region, considering the collision of a heavy particle in the weakly-bound subsystem of the 
remaining ones.  This approach was used to interpret the results
obtained with the renormalized zero-range model~\cite{2012frederico}, as well as with the Gaussian finite-range interactions.

To simplify our study, we assume no interaction between the heavy particles 
and that the heavy-light molecule ($HL$)  has a weakly-bound energy $B_2$. 
When $B_2\to 0$ the three-body Efimov levels are given by 
$B_{3}^{(n)} \to e^{-(2n\pi/s_0)} B_3$, where $ B_3\equiv B_3^{(0)}$ is the 
ground state binding energy of the models we use  in our approaches to obtain the  
$s-$wave cross-sections.

We start our analysis by introducing a scaling function for the dimensionless product of the $s-$wave cross-section 
and energy.  With $B_3$ and $B_2$ as the scales of the  $HHL$ system and $E$ the 
colliding energy at the three-body center-of-mass, this function can be written as
\begin{equation}
\sigma \,B_3\,={\mathcal S}\left(E/B_3\,, {B_2/B_3}\,, A 
\right),
\label{scalingf}
\end{equation}
where $A\equiv m_L/m_H$.
This is strictly valid at the zero-range limit where $B_2=1/(2 \mu_{HL} a_{HL}^2)$,
with $\mu_{HL}$ being the reduced mass for the $HL$ subsystem. Here and in the
next, the units are such that $\hbar=1$ and $m_L=1$. 

\begin{figure}[thb]
\begin{center}
\includegraphics[width=8cm]{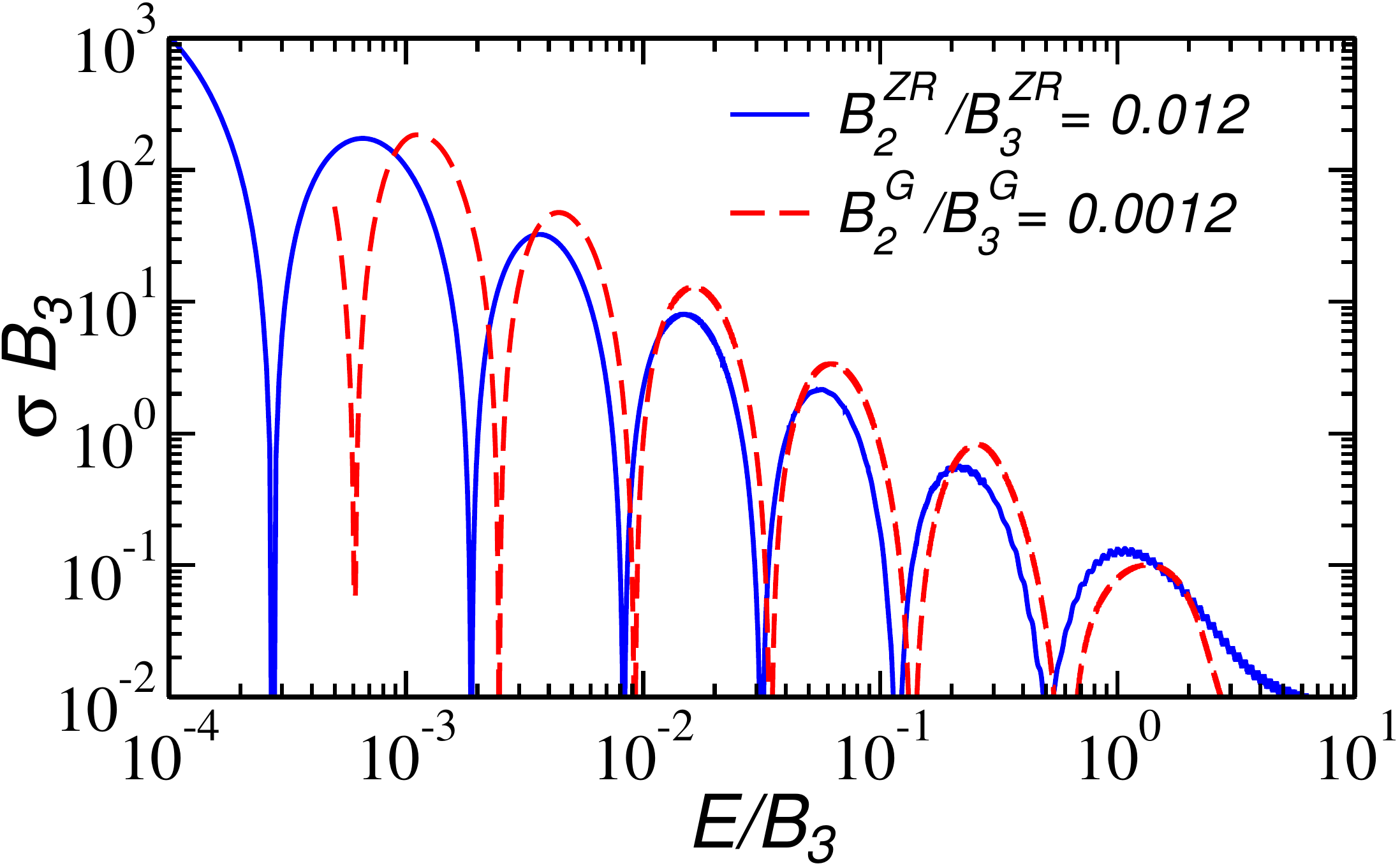}
\end{center}
\vspace{-0.6cm}
\caption{ 
The $s-$wave cross-section is shown as a function of the energy-collision $E$, for zero-ranged (ZR)
(blue-solid lines) and finite-ranged Gaussian (G) (red-dashed lines) potentials, for fixed mass ratio $A=
0.01$ and given two-body energies ($B_2^{G}$ is a factor smaller than $B_2^{ZR}$ to keep both 
results close to the unitary limit). Results are in units of $B_3$.
} 
\label{fig1}
\end{figure}

The scaling function for $A=0.01$ is shown in Fig. \ref{fig1} for the renormalized zero-range model \cite{2016shalchi} 
and for the Gaussian potential model calculated with the method developed in Ref.~\cite{BarPRL2009}, which was
extended to energies above the breakup threshold in Ref.~\cite{2012GarridoPRA86}. The 
Gaussian potential with $r_0$ being the interaction range is given by
\begin{equation}
V(r)=V_0 \, e^{-r^2/r_0^2},
\label{gauss}\end{equation}
where we have used $a_{HL}/r_0=50$ and $B_2/B_3=\,0.0012$. 

Noticeable, are the minima of the $s-$wave cross-section, where $k\cot\delta_0$ has poles. We observe that
positions of such poles tend to obey the Efimov law for $(k\,a_{HL})^{-1}\to 0 $. 
Between the zeros, there is a sequence of maxima for the cross-section
where the phase-shift passes through $(2n+1)\pi/2$, as seen in Fig. \ref{fig1}. 
It is tempting to associate the maxima obtained for the cross-section with resonances; however, a calculation 
by using the complex scaling method \cite{complex-scaling} for the Gaussian potential, excludes that.
These results are also corroborating the conclusions of \cite{2008yamashita,2017Deltuva} for the neutron-$^{19}$C 
system, where no resonance is found when changing the neutron separation energy in $^{19}$C.

\begin{figure}[thb]
\begin{center}
\includegraphics[scale=0.45]{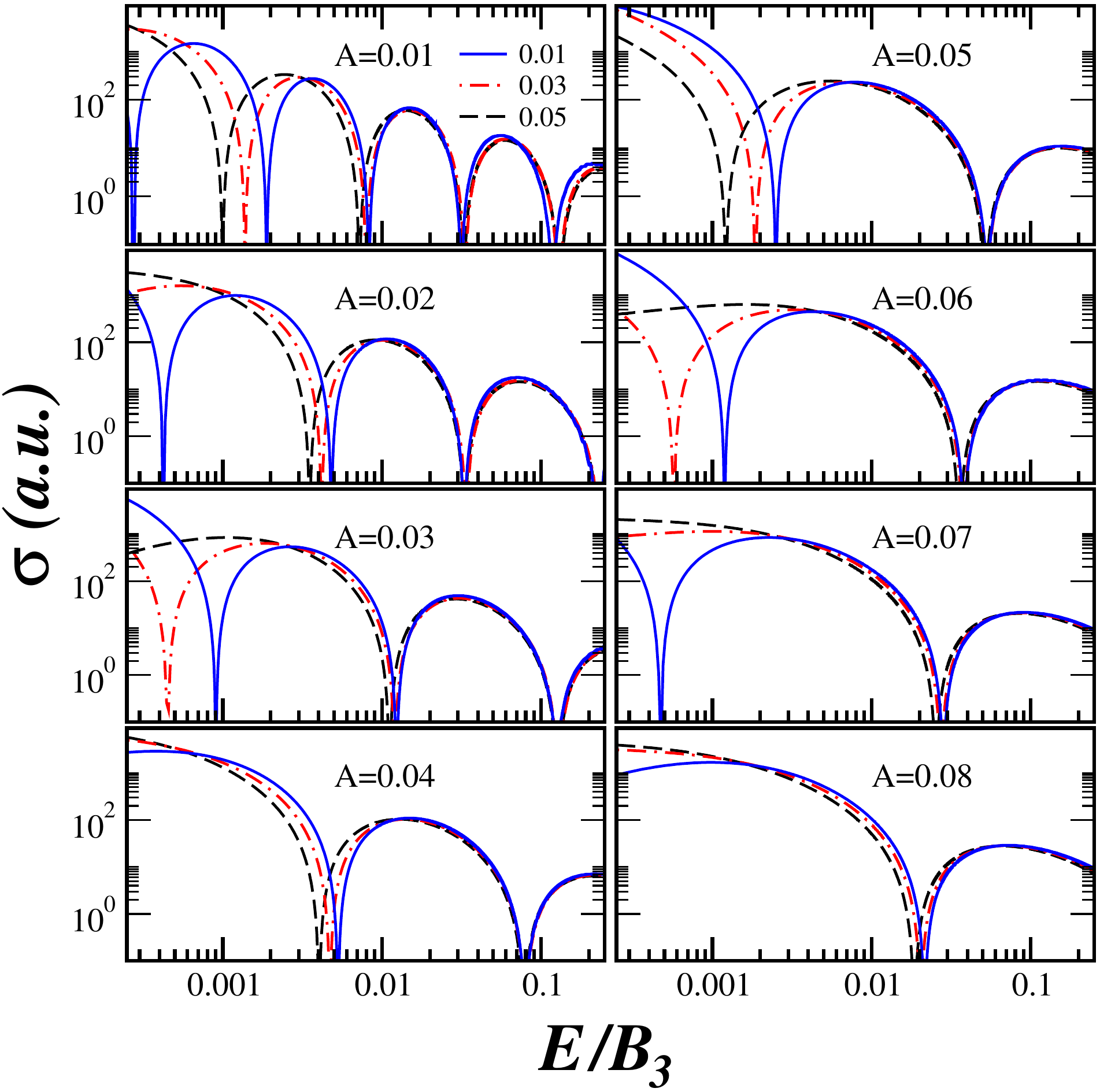}
\end{center}
\vspace{-.6cm}
\caption{The zero-ranged results for $\sigma$ (in arbitrary units) as functions of 
$E/{ B}_3$ are given in eight panels, with mass-ratios $A$ as shown inside the frames. 
In all the panels the two-body energies are fixed such that $B_2/{B}_3=$0.01 (solid-blue lines), 0.03 
(dot-dashed-red lines) and 0.05 (dashed-black lines).
}  \label{fig2}
\end{figure}

By considering different mass-ratios $A$, varying from 0.01 till 0.08, our
results for the cross-sections $\sigma$ (in arbitrary units)  are presented in Fig.~\ref{fig2} for 
three fixed weakly-bound two-body energies $B_2/{ B}_3=$0.01, 0.03 and 0.05. In the given  
eight panels we are presenting $\sigma$ as a function of $E/{ B}_3$.
From these panels, one can notice a sequence of zeros (or minima) which can appear for $\sigma$ 
as we decrease the mass ratio, by considering a fixed interval for the collision energy, such that
$E/{B}_3<1$.
Within this interval, when the mass-imbalance is less pronounced, e.g. for $A=0.08$, we can verify the 
occurrence of only one zero for $\sigma$ within the given energy range, whereas for  
$A=0.01$ it is possible to verify the existence of up to six zeros. 
Therefore, the large mass asymmetry ($A<<1$) is more favorable 
for the occurrence of several zeros/minima in $\sigma$. In order to verify the emergence
of a possible scaling factor between the position of successive zeros/minima in the $s-$wave cross-section, 
in correspondence with the Efimov bound-state spectrum, one should be able to extrapolate the 
two-body bound-state energies to the unitary limit (i.e., to $B_2=0$). 

Corresponding to the upper-left panel of Fig.~\ref{fig2}, when  $B_2/{ B}_3=0.01$ and
 A=0.01, we also have  Fig.~\ref{fig1} where $E/B_3$ was extended up to 1, which 
showed that it is possible to observe another minimum in $\sigma$ for a collision 
 energies much larger than the breakup threshold.  
 As we can observe, in this case, the value of the minimum in $\sigma$ is affected by 
 absorption, an expected behavior for energies above the break-up threshold. Therefore, $\sigma$ 
 is not being reduced to zero, but have just a minimum, with the value of the energy $E$
 also being deviated slightly to the right as $B_2$ is increased in Fig.~\ref{fig2}.
The ratio between the energy position of the successive zeros is about the Efimov geometric factor 
as one can easily check (we will explore such feature in a systematic  way later on), and as one could 
expect it should be  distorted by absorption effects, but far away from the breakup threshold. 

 It is noticeable to find minima of the cross-section for $E>>B_2$ and 
 quite deeply immersed in the three-body continuum, where still the
$s-$wave inelasticity parameter is very close to unity. This astonishing suppression of the breakup channel 
for energies of about two orders of magnitude the two-body binding is a manifestation of the 
long-range coherence between the heavy and light particles and the associated diluteness of the 
target, making it hard to destroy the system, where the light particle binds with any one of the 
heavy particles and the dynamics is dominated only by the exchange of the light particle between the 
two heavy ones. The $HL$ molecule becomes invisible to the collision of the heavy one. 
Semi-classically, the possibility of the destructive interference between the direct trajectory 
and the one from the exchange process gives the zeros of the phase-shift. 

The fact that the breakup channel is suppressed is closely related to the non-existence of resonances. 
In the adiabatic hyper-spherical representation of this mass imbalanced three-body system, it happens that
the coupling between the lowest adiabatic channel, which asymptotically goes to the
atom-dimer channel, with the breakup channels is weak (see e.g. \cite{2012GarridoPRA86}). 
In addition,  asymptotically the lowest
adiabatic hyper-spherical potential is attractive, while the breakup channels have a barrier around
$\rho\sim |a_{HL}|$. Indeed, in the case of Borromean systems, such barrier makes the Efimov turn to a 
continuum resonance when  $|a_{HL}|$ is decreased \cite{BringasPRA04}.
 	
 We summarize the findings presented in figures \ref{fig1} and \ref{fig2} as: 
 (i) the number of minima of  the $s-$wave cross-section decreases significantly when $A$ and 
 the Efimov ratio increases, and (ii) more minima are seen when  $B_2/B_3$ decreases. Particularly, with respect to 
 the second point, we found that the zeros of the cross-section are coming out from the scattering threshold and the
 $H-(HL)$ scattering length passes through zero values  
 when  $B_2/B_3$ is driven towards the more favorable condition for the  Efimov effect. 
 That is the counterpart of the unitary limit where virtual states come from the second energy sheet 
 to become bound states. In the continuum region, zeros and maxima of the cross-section come one by one as
$B_2/{ B}_3\to 0$, which completes the final picture of the Efimov limit including the scattering region. 

The manifestation of the Efimov discrete scaling in the atom-molecule collision can be
systematically studied by the ratio between the energies of successive zeros/minima as a function of the mass ratio and
a dimensionless ratio between two and three-body scales as follows.
For that, a scaling function is introduced relating the energies of two adjacent minima obtained for the 
cross-section $\sigma$. Within a convention that $E_{n+1}>E_n$, this function is given by
\begin{equation}\label{ratio}
E_{n+1}/E_n={\cal R}\left( 1/(E_{n+1}^ {1/2}\, a_{HL}), A \right),
\end{equation} 
where ${\cal R}\left( 0, A\right)=\text{e}^{2\pi/s_0}$ is the unitary limit.

\begin{figure}[thb]
\begin{center}
\includegraphics[scale=0.4]{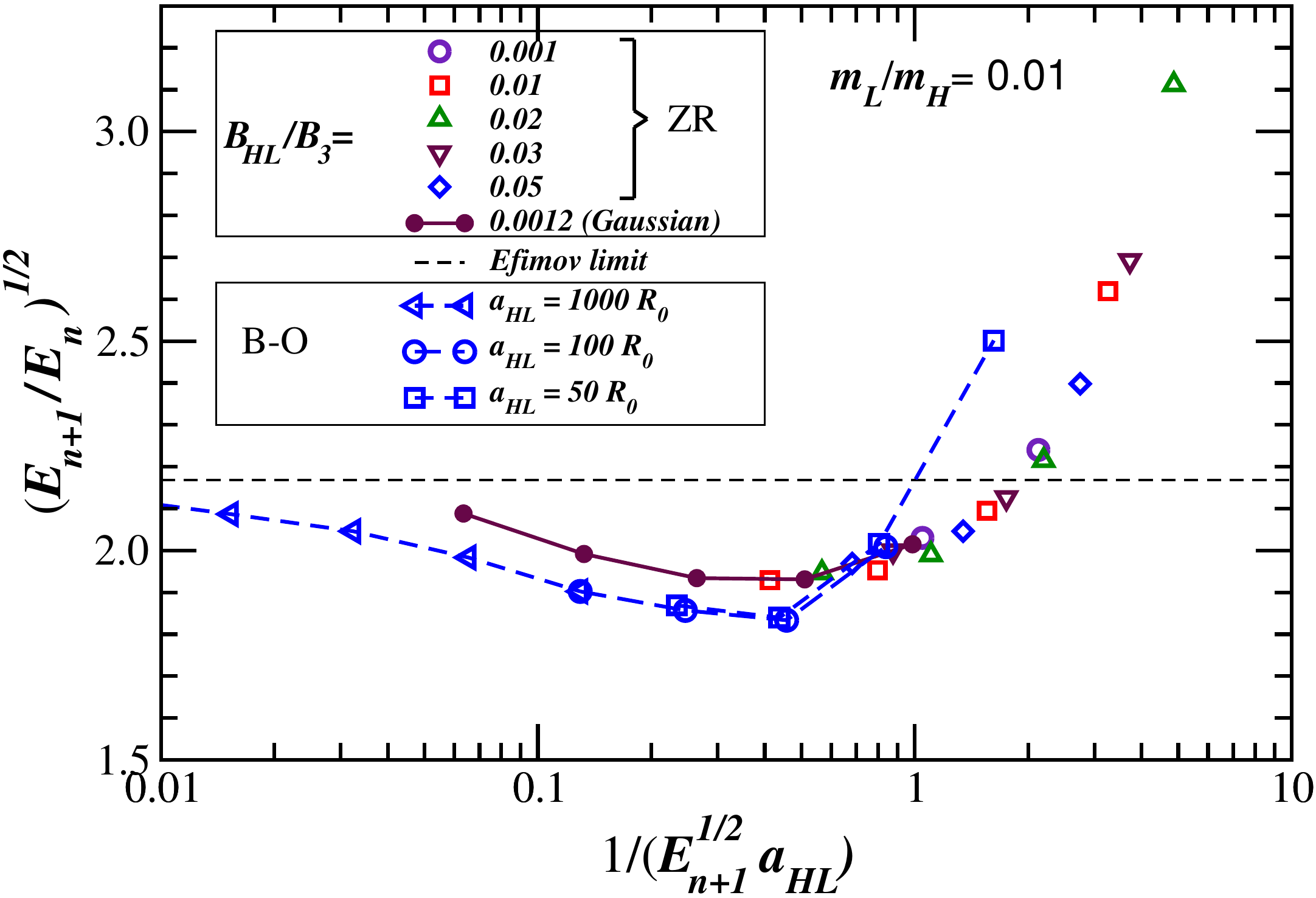}
\end{center}
\caption{Ratio between the energy positions of  the successive zeros ($n+1, n$) of the cross-section $\sigma$  plotted versus 
$1/(E_{n+1}^{1/2}\, a_{HL})$ for mass-ratio $A=0.01$. 
The results obtained with renormalized zero-range (ZR) model, for given two-body energies, are indicated inside the frame.
The straight dashed line indicates the Efimov limit for $A=0.01$.
The solid-line with dots shows the results obtained with the Gaussian potential (\ref{gauss}).
The Born-Oppenheimer (B-O) results (connected by a dashed-blue line) are shown for three boundary
conditions. 
}
\label{fig3}\end{figure}

The universal scaling function (\ref{ratio})  is shown in Fig. \ref{fig3} for the extreme case
$ A=0.01$, calculated with the Gaussian and zero range potentials. The curious behavior 
of the scaling function around the Efimov ratio, indicated by the horizontal dashed line, 
by departing from the unitary limit decreases, has a minimum and then increases, namely the zeros 
more distant. Note that we have plotted results for the renormalized zero-range potential with different $B_2$ and 
scattering lengths, ranging from $0.001 \leq B_2/B_3 \leq 0.05$. With the Gaussian potentials, within our numerical accuracy,
we were able to approach more closely the Efimov limit. However, when going to smaller values of $1/(E_{n+1}^{1/2}\, a_{HL})$,
we stand above the breakup threshold and evidently the coupling to  the breakup channel affects the ratio, as the figure suggests.  

The curious behavior of the ratio,  namely when starting from smaller to 
larger collision energies it is first above  the Efimov geometrical factor then it decreases and increases again
towards it, can be qualitatively understood by considering the collision within the BO approximation. In this case,
the effective $H-H$ long-range potential is supplemented by a boundary condition at some short distance $R$, 
with the continuity of the logarithmic derivative of the wave-function $u(R)$ imposed at $R=a_{HL}$. 
In our illustration, the elastic scattering $S$-matrix is found from the boundary condition at $R=a_{HL}$.
To make our point clear, we assume no two-body $H-H$ potential; and, we expand the effective BO potential~\cite{1979fonseca}, 
where the leading-order term is $\sim 1/R^2$, and we also consider the effect of the next order term, 
implying in the inclusion of a Coulomb-like $1/R$ interaction.
Therefore, as one can extract from the expansion of the potential presented in \cite{1979fonseca}, 
we have the following effective two-body equation for the collision of the heavy particle $H$ with relative  momentum $k$ 
with respect to the $HL$ dimer:
\begin{eqnarray}
&&\left[-\frac{d^2}{dR^2} - \frac{s_0^2+\frac{1}{4}}{R^2}g\left(\frac{R}{a_{HL}}\right)
\right] u(R)= k^2 u(R), 
\label{sr5}
\end{eqnarray}
where $g(y) \equiv 1+2 y+2.07 y^2$, such that the leading term in the interaction,
$-\left({s_0^2+{1}/{4}}\right)/{R^2}$, provides the Efimov limit. The wave number is related to
the collision energy by $k=\sqrt{2 \mu_{H,HL}E}$, where 
$\mu_{H,HL}\equiv m_H (1+A)/(2+A)$.
The expansion for $g(y)$ is found  by requiring an approximation of the BO potential 
valid not only for $R\ll a_{HL}$, but also for $R/a_{HL}\sim 1$.
With this approximation, the Coulomb-like correction
$-2\left({s_0^2+{1}/{4}}\right)/{(a_{HL}R)}$ is added to the Efimov term, as well as a constant which is
negligible for larger scattering lengths.
As shown by \cite{1979fonseca}, in case of negative-energies we can obtain exact solutions 
for the Eq.~(\ref{sr5}), given by Bessel functions in case we consider the leading term $1/R^2$ 
for the interaction.  In the present extension to scattering energies, we can also verify analytical solutions 
for the Eq.~(\ref{sr5}), which are given by Whittaker functions.
This eigenvalue equation has no lower bound energy, namely, the Thomas collapse is present, 
which requires a short-range scale imposed by a boundary condition at $R=R_0\ll a_{HL}$. 
In what follows, a hard wall will be used, and from the boundary condition at $R=a_{HL}$ the phase-shift 
is finally obtained. In this way, the log-periodicity of the $s-$wave phase-shift with the 
energy is only deformed by the presence of the $1/R$ contribution.

As a result, if  the BO potential in Eq. (\ref{sr5}) is given only by the Efimov term, the ratio (not plotted in Fig.~\ref{fig3}) 
would approach the Efimov limit monotonically from above when decreasing $1/(E_{n+1}^{1/2}\, a_{HL})$. 
The minimum observed in the BO results (dashed-blue curve in Fig.~\ref{fig3}) comes from the Coulomb-like correction. 
As shown by using different values for the position of the hard wall at short distances, there are no significative  
range corrections. 
Therefore, we note that the first two terms of the BO potential are quite relevant to provide a qualitative description of 
the scaling function. 
This approximation is working surprisingly well in particular for large values of $E_{n+1}^{1/2}\, a_{HL}$, when 
approaching the Efimov limit, considering that in this limit the coupling to the breakup channel (which is not being
taken into account) is expected to be relevant.
For smaller values of $E_{n+1}^{1/2}\, a_{HL}$ the expansion of the BO potential starts to 
breakdown
due to its poor efficacy when decreasing the collision energy, with the wavelength being of the order of the scattering 
length.

{\it Practical implications.}
The poles of $k\cot\delta_0$, which correspond to the zeros/minima of the $s-$wave cross section,
are directly connected with the Efimov spectrum of the heavy-heavy-light $(HHL)$ system near 
the unitary limit. This is shown by considering a mass-imbalanced system $A<<1$ with no interaction 
between the two-heavy particles and with the heavy-light sub-system bound with energy close to zero (near unitary limit). 

By considering the mass ratio between Li and Yb, $A=$0.034, the cross-section for the Yb + LiYb collision
can in principle present a couple of zeros. 
We can imagine a situation where $a_{\rm {YB-Li}}$ is adjusted at some large positive values, with the colliding energy
being varied slowly. In this case, $\sigma$ should 
present minima at some specific colliding energies, whose positions are approximately geometrically spaced.
In conclusion, we suggest as the best possible situation to probe the Efimov discrete scaling in the continuum to consider the
atom-molecule scattering with large mass asymmetry through cold collisions, which are now feasible
\cite{2006krems}. The challenge in these experiments would be to control the scattering length towards the large 
values and then observe the cross-section minima at geometrically spaced colliding energies.

\noindent {\it Acknowledgements:} {\small
The authors acknowledge partial support from 
Conselho Nacional de Desenvolvimento Cient\'\i fico e Tecnol\'ogico
[Processes 308486/2015-3(TF), 302075/2016-0(MTY), 306191-2014-8(LT)], 
Funda\c c\~ao de Amparo \`a Pesquisa do Estado de S\~ao Paulo
[Projects 2017/05660-0(LT and TF),  2016/01816-2(MTY) and 2013/26258-4(TF)],
Coordena\c c\~ao de Aperfei\c coamento de Pessoal de N\'\i vel Superior
[Proc. 8888.1.030363/2013-01 (MAS and MTY) and Senior visitor program in ITA-DCTA (LT)].
National Science Foundation [Contract No. NSF-HRD-1436702 with Central State University (MRH)]
and the Institute of Nuclear and Particle Physics at Ohio University (MRH).
Spanish Ministerio de Economia y Competitividad [contract FIS2014-51971-P] and 
Consejo Superior de Investigaciones Cient\'\i ficas [project i-LINK 1056] (EG).}

\end{document}